\documentclass{optica-article}

\journal{opticajournal} 

\articletype{Research Article}

\usepackage{lineno}
\begin{document}

\title{High-throughput speckle spectrometers based on multifractal scattering media}

\author{Bhupesh Kumar,\authormark{1,$\dagger$} Yilin Zhu,\authormark{2,$\dagger$}, Luca Dal Negro,\authormark{2,3,4,*} and Sebastian A. Schulz\authormark{1,**} } 
\address{\authormark{1}School of Physics and Astronomy, University of St Andrews, North Haugh, St Andrews KY16 9SS, UK\\
\authormark{2}Division of Material Science and Engineering, Boston University, 15 Saint Mary's Street, Brookline, Massachusetts 02446, USA\\
\authormark{3}Department of Electrical \& Computer Engineering and Photonics Center, Boston University, 8 Saint Mary's Street, Boston, Massachusetts 02215, USA\\
\authormark{4}Department of Physics, Boston University, 590 Commonwealth Avenue, Boston, Massachusetts 02215, USA\\
\authormark{$\dagger$}These authors contributed equally to the work. 
} 

\email{\authormark{*}dalnegro@bu.edu\\\authormark{**}sas35@st-andrews.ac.uk} 

\begin{abstract*} 
We present compact integrated speckle spectrometers based on monofractal and multifractal scattering media in a silicon-on-insulator platform. Through both numerical and experimental studies we demonstrate enhanced optical throughput, and hence signal-to-noise ratio, for a number of random structures with tailored multifractal geometries without affecting the spectral decay of the speckle correlation functions. Moreover, we show that the developed multifractal media outperform traditional scattering spectrometers based on uniform random distributions of scattering centers. Our findings establish the potential of low-density random media with multifractal correlations for integrated on-chip applications beyond what is possible with uncorrelated random disorder. 
\end{abstract*}
\section{Introduction}
In recent years, multi-functional disordered photonic devices have attracted significant
interest in photonics and nano-optics technologies \cite{Cao2022APR,Coluccelli2016NatComm,FarajiDana2018NatComm,Hartmann2020AOM,Qiao2022ACSPhotonics,Redding2013NatPhotonics,Redding2014Optica,Redding2016Optica,Wang2019NatComm,Xiong2022Optica,Yang2021Science}. In particular, speckle-based spectrometers operating in the multiple scattering regime of disordered random media overcome the fundamental trade-off between size and spectral resolution, resulting in record-high performances as well as noise robustness \cite{Redding2013NatPhotonics}.
Multiple scattering of light when propagating through strongly scattering random media leads to an
increased optical path lengths with a wide spectral path length spread that provides high spectral resolution at small footprints. 
However, most disordered structures investigated in this context consist of uncorrelated random patterns or amorphous photonic systems with a limited number of spatial correlations and necessarily feature a large and approximately uniform density of scattering particles. This leads to a very diffuse propagation regime, multiple-scattering induced reflection \cite{Cao2022APR}, increased out-of-plane scattering from etched hole sidewalls and surface state absorption, severely limiting their optical throughput. Therefore, there is presently a need to develop and explore alternative approaches to speckle-based on-chip spectrometers with tailored disorder enabling strongly scattering multi-scale random structures with significantly reduced density of scattering centers compared to traditional random systems. 

In order to address this challenge, in the present work we present on-chip spectrometers based on speckle correlations in low-density, non-homogeneous multifractal arrays with varying degrees of structural correlations created by random multiplicative cascade processes as described in \cite{Chen2023PRB}. 
Different from traditional monofractal systems, multifractal structures are characterized by multi-scale local density fluctuations described by a continuous function, i.e., their multifractal spectrum $f(\alpha)$, as opposed to a single scaling exponent (i.e., the fractal dimension) \cite{Chen2023PRB,Frisch1980Annals,nakayama2003fractal,falconer2004fractal,mandelbrot1982fractal,harte2001multifractals,martinez1990clustering,anitas2019small}. Therefore, compared to monofractals, amorphous, or traditional random media, multifractal point-patterns necessarily feature broader distributions of particle separations and reduced array densities, making them ideally suited for the realization of miniaturized high-throughput random spectrometers. 

\begin{figure}[ht!]
	\begin{center}
		\includegraphics[width=12cm]{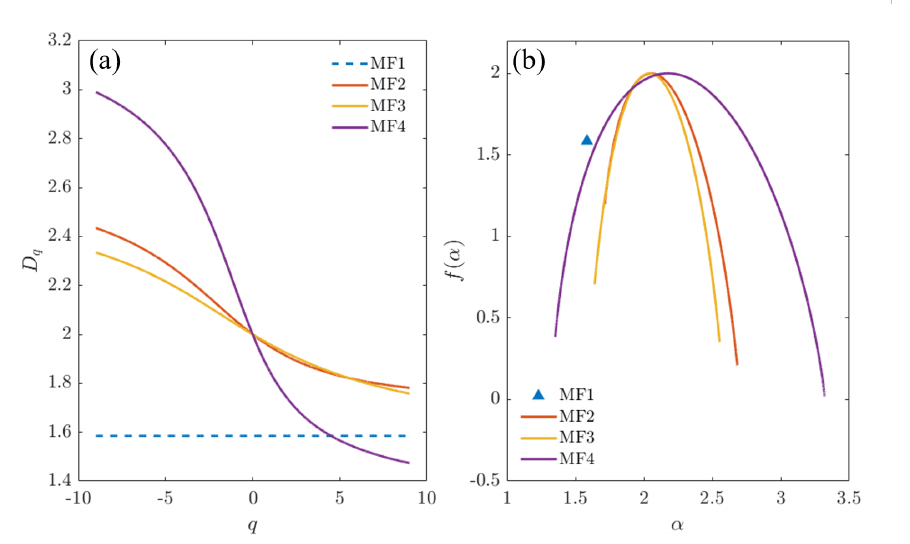}
		\caption{(a) Spectra of generalized dimensions $D_q$ and (b) Multifractal singularity spectra $f(\alpha)$ for multifractal structures generated with probability vectors $p=[1,1,1,0]$ (`MF1', dashed blue and triangle symbol), $p=[1,1,0.75,0.5]$ (`MF2', solid red), $p=[1,0.75,0.75,0.5]$ (`MF3', solid yellow), and $p=[1,0.75,0.5,0.25]$ (`MF4', solid purple).}\label{fig:MFspectra}
	\end{center}
\end{figure}

Multifractal point patterns with different degrees of structural correlations are constructed from the non-Gaussian probability fields generated by the random multiplicative cascade model \cite{anitas2019small} with the initial probability vectors $p_{i}\in[0, 1]$ with $i = 1, 2, 3, 4$ in conjunction with a Monte Carlo rejection scheme, as detailed in \cite{Chen2023PRB,martinez1990clustering,anitas2019small}. Specifically, we designed devices with four representative types of tailored random structures ranging from monofractals to multifractals with increasing degree of structural inhomogeneity. 
In Figure 1(a) we show the spectra of generalized fractal dimensions $D_q$ computed from the general expression \cite{martinez1990clustering}:
\begin{equation}
    D_q=\frac{1}{1-q}\log_2{(f_{1}^{q}+f_{2}^{q}+f_{3}^{q}+f_{4}^{q})}
\end{equation}
where $f_{i}=p_{i}/\sum_{i=1}^{4}p_{i}$ with $i=1,2,3,4$. The corresponding
multifractal spectrum $f(\alpha)$, shown in Figure 1(b), is computed by the Legendre transform method as detailed in \cite{Chen2023PRB,martinez1990clustering,anitas2019small}.
The computed spectra in Figure 1(a,b) show a clear transition from a homogeneous fractal
structure with constant box-counting $D_0\approx 1.58$ and
$f(\alpha)$ supported only by a single point, to more inhomogneous
multifractals arrays characterized by increasingly broader spectra
when the amplitudes of the initial probability vectors are decreased.

As discussed in the next section, these patterns are fabricated on a silicon-on-insulator wafer and the resulting device structures investigated in the near-infrared (NIR) wavelength range. 
\begin{figure}[ht!]
\begin{center}
\includegraphics[width=12cm]{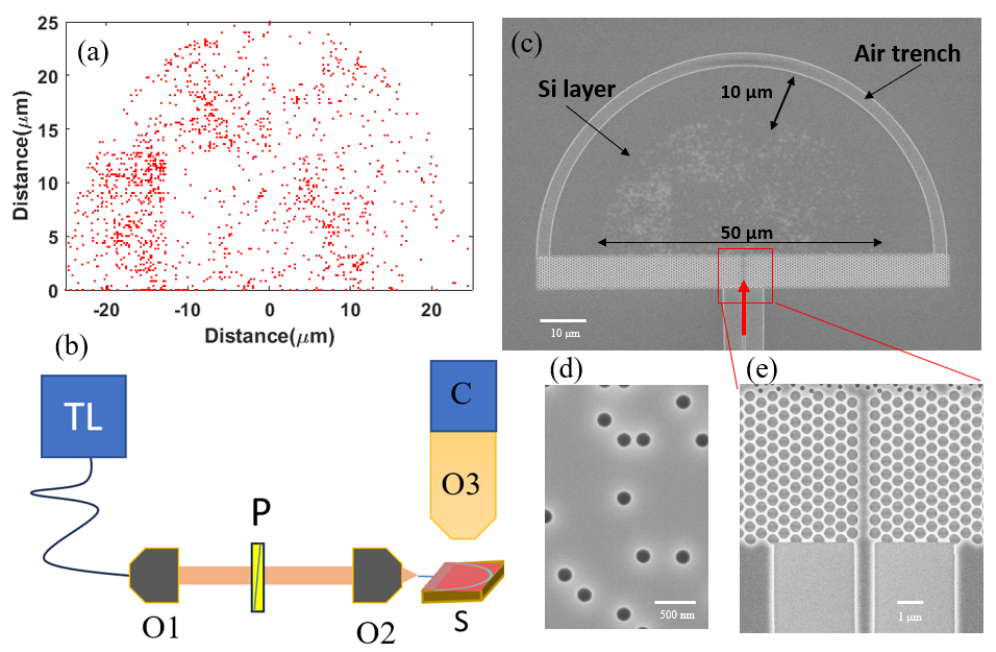}
\caption{(a) Multifractal point pattern MF4, generated with probabilities p= [ 1.00, 0.75, 0.50, 0.25]. (b) Schematics of the experiment setup. TL: NIR Tunable laser, wavelength range: 1500 nm-1630 nm, O1:  Collimating objective, O2: focusing objective, P: Polarizer, O3: 50X NIR microscope objective, C: NIR camera, S: sample.  (c) Scanning electron microscope image of the fabricated device on a Silicon on Insulator (SOI) wafer. The scatterers (air holes, Diameter= 150 nm) are distributed as a multifractal pattern distribution (in this image the distribution MF4 shown in panel a) in a semicircular area of radius 25 $\mu$m.  A Photonic crystal mirror is fabricated at the base to avoid light loss due to back reflections. An outer air trench of width 2 $\mu$m is used as a scattering edge for the transmitted light. The component scattered out of plane at this edge is then collected using the NIR objective. The air trench and scatterer distribution are separated by a 10 $\mu$m silicon slab propagation region, to ensure a clear delineation between light scattered by the air trench and out-of-plane scattering from the multi-fractal region. The insets in the bottom row (d,e) are magnified sections of the SEM image, showing the photonic crystal mirror, the waveguide serving as the spectrometer input and the representative shape of the air holes.}\label{fig:set_up}
\end{center}
\end{figure}
\section{Sample fabrication and Experimental Set-up}\label{sec:experiment}
To experimentally compare the performance of the different mono- and multifractals described above as speckle spectrometers we fabricated identical spectrometer layouts incorporating the four different fractals and also one uniform random scatterer distribution. For this purpose, the multifractal algorithm, together with the aforementioned probabilities, is used to generate a point distribution in a semicircle with a radius of 25 $\mu$m. Each point in the distribution then becomes the centre coordinates for an exposure region during electron beam lithography, using the positive tone AR-P 6200 resist. The exposure is then followed by pattern transfer into the top silicon layer of a 220 nm Si on 3 $\mu$m silica silicon-on-insulator wafer using an SF6/CHF3 reactive ion etch, such that the air holes created have a radius of 150 nm and a minimum separation of 50 nm. As shown in Figure \ref{fig:set_up} the spectrometer also features a photonic-crystal boundary in the bottom to suppress reflection, an outer air trench to scatter the transmitted speckle pattern for collection using a vertically mounted objective and NIR camera, and incoupling waveguides. All these elements are defined during the same electron-beam lithography
exposure as the scatterer distributions.
A scanning electron microscope (SEM) image of the
device is shown in Figure \ref{fig:set_up}.
The photonic crystal (PhC) mirror features a complete bandgap in the wavelength region of interest (1500-1525 nm) consists of a triangular lattice of airholes with a lattice period of $a = 505$ nm and a hole radius of 180 nm. A W1 defect waveguide in the center of the PhC allows the incoming light to couple into the spectrometer. A semi-circular air trench of width  2$\mu$m at a radial distance of 10 $\mu$m from the outer edge of the semicircular scattering area is used to collect light for detection.
The devices were characterized using a tunable, near-infrared laser(NIR) (TS-100, EXFO), coupled to a single-mode, polarization-maintaining fiber. The fibre output is collimated using an aspheric lense, passed through the polarizer to get the pure TE component of the incoming light and then focused onto the cleaved edge of the access waveguide (4 $\mu$m width) using a second aspheric lense.
The light Scattered from the device was imaged from above using a 50X objective (Mitutoyo  NIR Infinity Corrected) and an InGaAs camera (Raptor, Owl 640 II). All the experiments are performed at room temperature on a floated optical table.

\begin{figure}[ht!]
\centering\includegraphics[width=8cm]{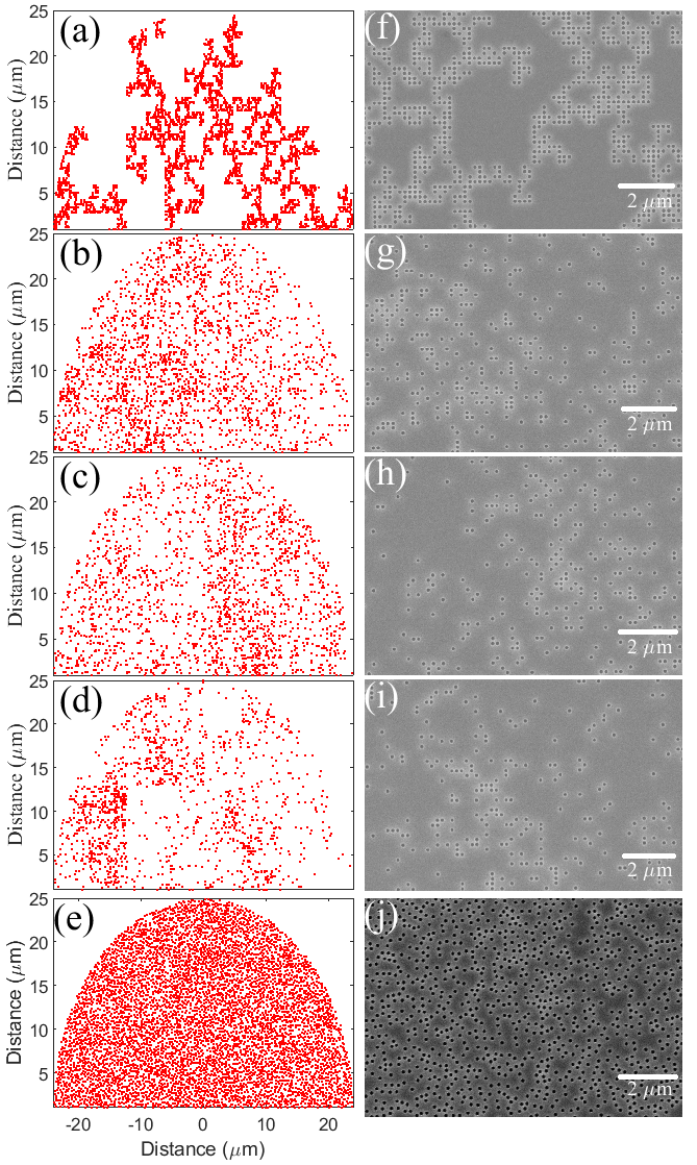}
\centering\caption{a,b,c,d are the real space point patterns for MF1-MF4. (e) Shows real space point patterns for the uniform random (UR) distribution. ( f, g, h, i) are SEM images of sections of MF1 to MF4 respectively and (j) shows the SEM image of a zoomed section of the UR distribution.}\label{fig:devices}
\end{figure}
\begin{figure}[ht!]
\centering\includegraphics[width=9cm]{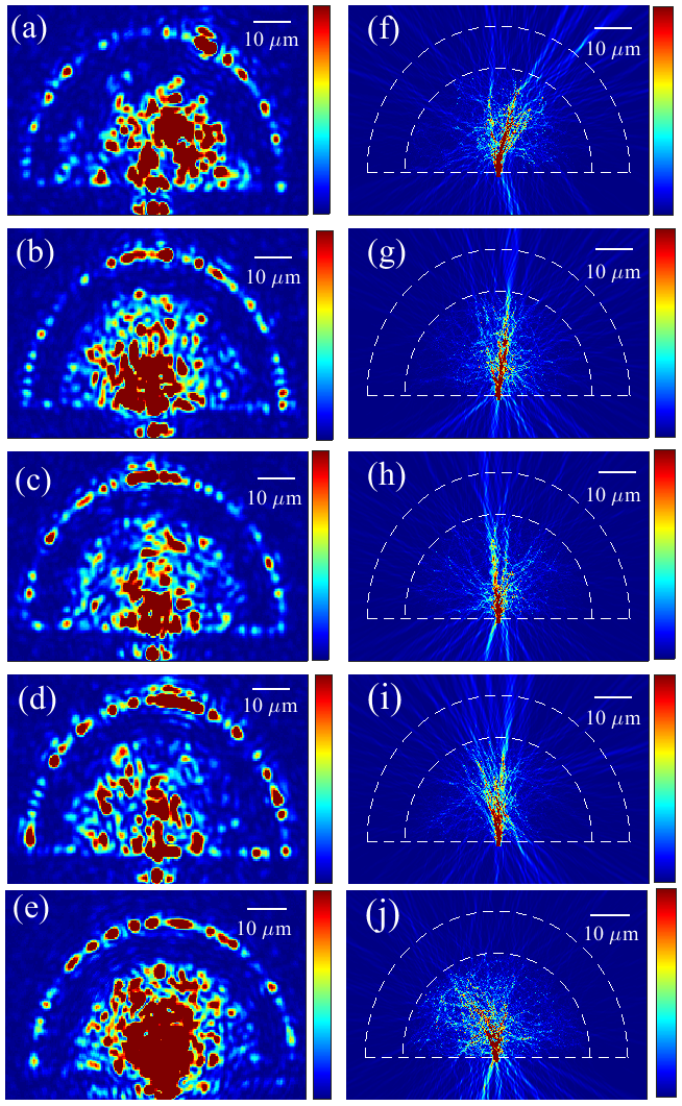}
\centering\caption{ Out-of-plane scattering images for spectrometers based on the point patterns MF1 (a), MF2 (b), MF3 (c), Mf4 (d) and UR (e) for an input wavelength $\lambda =1510$ nm. (f,g,h, i,j) show 2D generalized Mie theory-based simulations of the in-plane distribution of the corresponding pattern distributions at the same wavelength.}\label{fig:fieldimages}
\end{figure}
\section{Results}
To characterize the fabricated devices, the light is injected from the origin of the semicircle via a waveguide defect created in the PhC mirror. Due to multiple scattering light disperses throughout the scattering medium, before reaching the outer air trench. The non-uniform intensity (speckle) distribution along the air trench is highly wavelength-dependent and can be used as a unique fingerprint for the incoming light. Each device is scanned over a bandwidth of 15 nm from 1510 nm to 1525 nm with a step size of 0.2 nm. The intensity pattern corresponding to each wavelength is imaged from the top by a 50x objective and InGaAs IR camera. Figure \ref{fig:fieldimages}(a-e) shows the imaged intensity patterns recorded at a wavelength of 1510 nm for all the devices under investigation. To compare the performance of the different speckle spectrometers we take the following steps. At each wavelength the intensity distribution along the air trench is split into 25 equal parts, creating a [25 x 75] transmission matrix (TM). From this TM we can integrate the total power per wavelength (column) to extract the wavelength-dependent throughput of each spectrometer and approximate the spectral resolution via the speckle correlation function \cite{Cao2022APR,Redding2013NatPhotonics}:

\begin{equation}\label{eq:xcorr}
C(\lambda, D) = \frac{\langle I(\lambda,D)I(\lambda+\delta\lambda,D) \rangle } {\langle 
 I(\lambda,D) \rangle  \langle  I(\lambda+\delta\lambda,D) \rangle}-1,
\end{equation}
where $I(\lambda, D_{i})$ denotes the light intensity at wavelength $\lambda$ and
detector $D_{i}$ (i = 1, 2, …., 25), and the average is taken over the
respective wavelength ranges. The $I(\lambda, D_{i})$ values are obtained directly from the TM. Finally, the values $C(\delta \lambda, D_{i})$ are
subsequently averaged over all 25 detectors and normalized to
the value at $\delta$$\lambda$ = 0 to yield the speckle spectrometer correlation
function $C$($\delta$$\lambda$).

To support our experiment observations with theoretical analysis, we further conducted full-field simulations of the in-plane scattering behavior of the fabricated devices using two-dimensional generalized Mie theory (2D-GMT) \cite{Zhu2023AO,DalNegro2021GMT}. The 2D GMT is a semi-analytical spectral method that rigorously solves Maxwell's equations for 2D geometries of arbitrary arrays of scattering cylinders by expanding the fields into a sum of cylindrical Bessel and Hankel functions up to a specified cutoff angular order $\ell_\mathrm{max}$, thus obtaining a linear transfer matrix equation for the unknown field expansion coefficients \cite{dal2022waves,Gagnon2015GMT}. Compared to mesh-based traditional methods such as the finite element methods (FEM) or the finite difference time domain (FDTD), the 2D-GMT is mesh-free and provides unmatched accuracy at significantly reduced computational cost \cite{Gagnon2015GMT,Zhu2023AO}. In our GMT simulations, we considered large systems of $N\sim 2000$ particles under TE polarization and used the material and geometrical parameters of the fabricated devices. To approximate the 3D nature of the fabricated devices in our 2D simulations, the bulk dispersion of the dielectric material is replaced by the effective index of the fundamental guided single mode of an unperturbed 3D heterostructure \cite{Qiu2002effective,Trojak2020APL,Trojak2021APL}, i.e., the SOI wafer without air holes. During our simulations, we selected the cutoff order to be $\ell_\mathrm{max}=3$ and verified that it provides sufficient accuracy. 
In Figure \ref{fig:fieldimages}(f-i) we show the results for the in-plane field intensity distributions scattered by the four types of multifractal structures and Figure \ref{fig:fieldimages}(j) we display the simulated intensity of the reference uniform random structure. 
The overall spatial distributions of the simulated scattered intensities demonstrate strong multiple scattering inside the devices and qualitatively agree with the measured intensity distributions reported in Figure \ref{fig:fieldimages}(a-e).

\begin{figure}[ht!]
\centering\includegraphics[width=10cm]{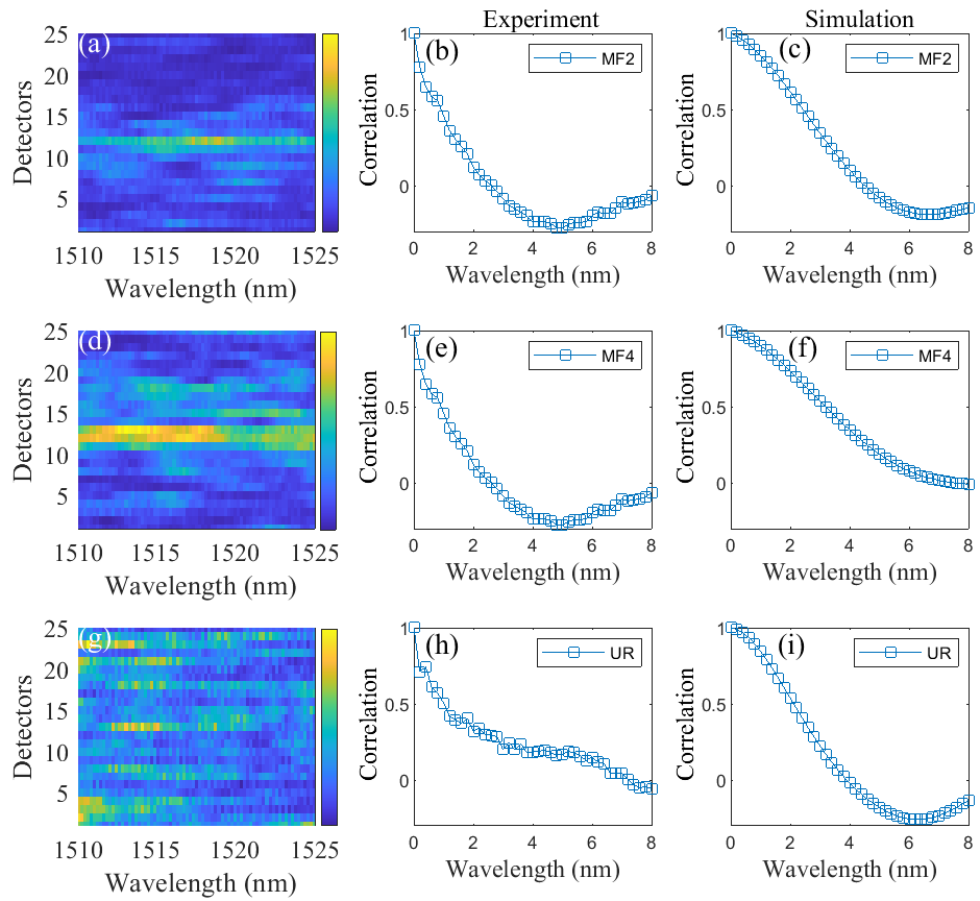}
\caption{(a,d,g) are  Transmission matrices (TM) obtained experimentally for near IR light measurements in the wavelength 1510-1525 nm with a step size of 0.2 nm for MF2, MF4 and the UR patterns, respectively. The detectors on the y-axis represent the number of sections the outer air trench is divided. Each transmission matrix represents the intensity distribution per detector of different wavelengths separated by 0.2 nm. (b,c)  Experimental and 2D numerically calculated cross-correlation functions respectively for point pattern MF2.  (e,f) Experimental and 2D numerically calculated cross-correlation functions respectively for point pattern MF4. (h,i) Experimental and 2D numerically calculated cross-correlation functions respectively for the UR point pattern.}\label{fig:xcorr}
\end{figure}

The measured transmission matrix selected multifractals and the uniform random distribution-based spectrometers are shown in Figure \ref{fig:xcorr}(a,d,g). We also show the measured cross-correlation of the corresponding devices in Figure \ref{fig:xcorr}(b,c,h). Using the correlation curve half-width-half-maxima (HWHM), the spectral resolution of each spectrometer can be estimated. For all devices considered here, the HWHM is almost identical, in the range of 1-1.3 nm. In Figure \ref{fig:xcorr}(c,f,i) we show the cross-correlation results obtained from the transmission matrix computed using the 2D-GMT method. During simulations we scan the devices with the same bandwidth and step size of wavelengths as in the experiments and calculated the cross-correlation of the in-plane field intensities at the simulated air trench location at all wavelengths recorded by 25 detectors uniformly spaced along the air trench, using Eq. \ref{eq:xcorr}. We note that the simulated cross-correlations well-capture the salient features of the experimentally measured ones and confirm the behaviour that all devices are expected to have the same HWHM. We thus obtain our first key results, namely that MF patterns with widely different scatterer densities and inhomogeneities in the scatterer density perform equally well in terms of spectrometer resolution and are comparable to the benchmark UR distribution.

\begin{figure}[ht!]
\begin{center}
\centering\includegraphics[width=12cm]{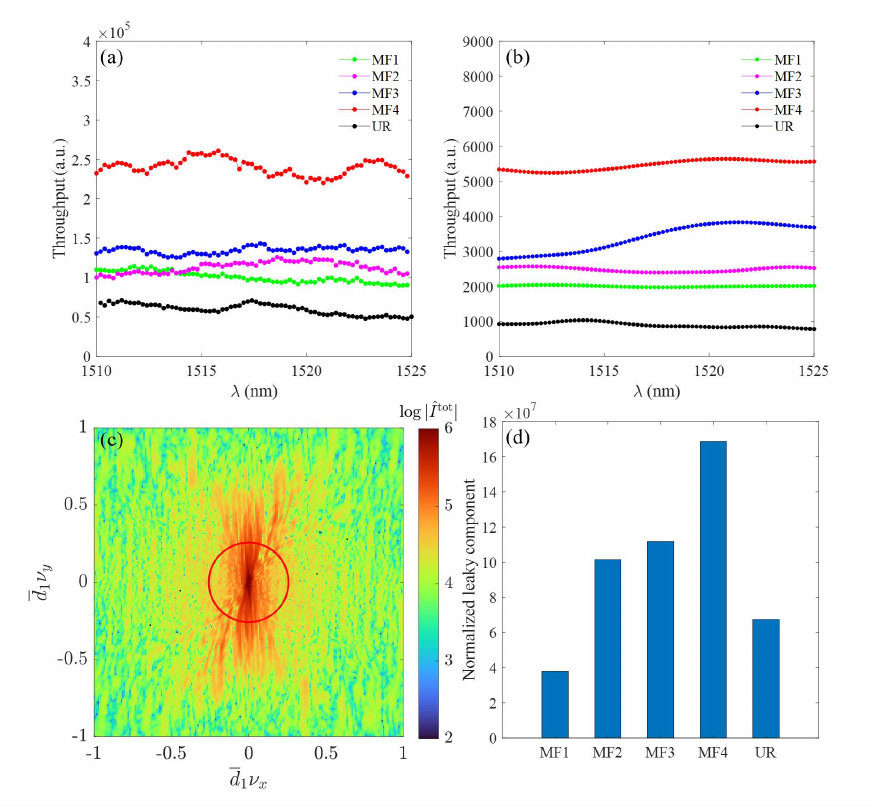}
\caption{(a) The integrated light scattered from the air trench of all samples including the UR distribution is plotted as a function of wavelength in the range 1510 nm to 1525 nm. (b) 2D-GMT simulated in-plane light integrated over the same air trench region, normalized by the number density of each structure. (c) 2D Fourier transform spectra of the electric field intensity distributions shown in Fig. \ref{fig:fieldimages} (i). Note the spatial frequency is normalized to averaged first neighbor distance $\overline{d}_1$ of scatterers. The area inside the red circle corresponds to the leaky region. (d) Integrated leaky component normalized by the number density of each structure, where `MF1' to `MF4' correspond to fractal structures in Fig. 3(a) to Fig. 3(d) and `UR' correspond to uniform random structure in Fig. 3(e).} \label{fig:throughput}
 \end{center}
\end{figure}

However, all spectrometer systems feature trade-offs between the varying performance metrics. In the context of speckle spectrometers the main trade-off is expected between the device throughput (a measure of the sensitivity and signal to noise ratio) and the resolution/speckle correlation \cite{Cao2022APR}. Therefore for a true comparison of the performance of the different spectrometers we need to not only compare the HWHM of the correlation function, but also the device throughput. As mentioned earlier the throughput of each devices as a function of wavelength is measured by integrating the light intensity across the outer air trench for each wavelength step. To ensure the accuracy in measurements for each scattering configuration, the throughput is averaged over 5 identical devices fabricated on the same chip.
Laser power, polarization, and imaging camera gain are kept constant for all the measurements and chosen such that there is no saturation of the camera during the measurement process.
Figure \ref{fig:throughput}(a) shows the measured optical throughput for all the devices. We can see that the throughput corresponding to the most inhomogeneous multifractal structure (MF4) generated from the probability vector p = [1, 0.75, 0.5, 0.25] is four times larger than the one of the sample with uniform random distribution. Moreover, the two devices show comparable HWHM, and therefore have similar spectral resolution. Figure \ref{fig:throughput}(b) shows the simulated throughput of each device normalized by the corresponding number density $\rho_N$, which demonstrates good agreement with the experimental data in Figure \ref{fig:throughput}(a). The simulated throughput is calculated by integrating the field intensity over the air trench at each incident wavelength. It is noted that the normalized simulated throughputs capture the inherent behavior of the investigated geometries and are stable with respect to averaging over different realizations of the disorder with $<30\%$ fluctuations in the averaged transmission intensities.   

To further investigate the connection between the in-plane simulated transmission and the out-of-plane measured intensity we conducted frequency-space analysis on the simulated scattered intensity by estimating the overlap of the 2D spatial Fourier components of the scattered field intensity with the out-of-plane extraction condition determined by the red line in Figure \ref{fig:throughput}(c). In particular, the amount of field intensity leaking out of the photonic structure is estimated by integrating the spatial frequency components that fall inside a circle of spectral diameter $1/\lambda$ \cite{boriskina2008optical,akahane2003high,song2005ultra,srinivasan2002momentum,yamilov2006self}, which is then compared for all devices. Figure \ref{fig:throughput}(d) show the result of leaky component comparison. After normalizing the integrated leaky components by $\rho_N$, we find the counterintuitive result that the most inhomogeneous multifractal structure MF4 (shown in Figure \ref{fig:devices}(d)) leads to the highest out-of-plane scattering losses whilst also featuring the highest throughput. This indicates that the device throughput must be limited by other loss channels, such as reflections (even in the presence of the PhC mirror), light localization or surface state absorption \cite{Iadanza2020}. 

\section{Conclusion}
In this paper we designed and demonstrated compact speckle spectrometers based on tailored multifractal scattering media integrated on an SOI platform and compared their performances with uniform random media. Specifically, we considered structures generated by multiplicative cascade processes with tunable degrees of spatial homogeneity described by their multifractal singularity spectra. Using electron beam lithography we fabricated photonic devices consisting of circular air holes of diameter 150 nm and edge-to-edge separation 50 nm with high precision. By performing numerical simulations using 2D-GMT analysis of large scattering structures with the same parameters utilized in the experiments we demonstrated $4\times$ enhancement of the optical throughput of multifractals compared to uniform random systems, with similar spectral correlation characteristics. Moreover, we show that among the different multifractal media investigated, the most inhomogeneous one achieves the best throughput due to the lowest density of scattering holes, without loss of resolution. Our findings establish an effective approach to harness the multiscale nature of fractal structures for the engineering of more efficient speckle spectrometers of interest to lab-on-a-chip systems and miniaturized, cost-effective hyperspectral imaging devices.

\section*{Acknowledgments}
 S. A. Schulz and B. Kumar acknowledge funding from the EPSRC project  EP/V029975/1: "Disorder enhanced on-chip spectrometers". L.D.N. acknowledges support from the National Science
Foundation (ECCS-2015700, ECCS-2110204)"

\section*{Disclosures}
 ``The authors declare no conflicts of interest.''
 
\section*{Data availability} 
Upon publication the underlying data will be made available through the University of St Andrews repository. NOTE FOR REVIEWERS and EDITORS: after acceptance a doi and URL will be inserted here.




\bibliography{sample}

\end{document}